\documentclass[aps,prd,onecolumn,groupedaddress,showpacs,nofootinbib,amssymb]{revtex4}
\usepackage{graphicx,bm}
\usepackage[english]{babel}
\usepackage{amsmath}
\usepackage{amssymb}
\usepackage{amsfonts}

\begin{document}

\def\cL{{\cal L}}
\def\be{\begin{equation}}
\def\ee{\end{equation}}
\def\bea{\begin{eqnarray}}
\def\eea{\end{eqnarray}}
\def\beq{\begin{eqnarray}}
\def\eeq{\end{eqnarray}}
\def\tr{{\rm tr}\, }
\def\nn{\nonumber \\}
\def\e{{\rm e}}

\title{Covariant Lagrange multiplier constrained higher derivative gravity \\
with scalar projectors}

\author{Josef Kluso\v{n}$^1$, Shin'ichi Nojiri$^{2,3}$
and Sergei D. Odintsov$^4$\footnote{Also at Tomsk State Pedagogical
University, Tomsk, Russia}}

\affiliation{
$^1$ Department of Theoretical Physics and Astrophysics,
Faculty of Science, Masaryk University,
Kotl\'{a}\v{r}sk\'{a} 2, 611 37, Brno, Czech Republic \\
$^2$ Department of Physics, Nagoya University, Nagoya
464-8602, Japan \\
$^3$ Kobayashi-Maskawa Institute for the Origin of Particles and
the Universe, Nagoya University, Nagoya 464-8602, Japan \\
$^4$Instituci\`{o} Catalana de Recerca i Estudis Avan\c{c}ats
(ICREA) and Institut de Ciencies de l'Espai (IEEC-CSIC), Campus
UAB, Facultat de Ciencies, Torre C5-Par-2a pl, E-08193 Bellaterra
(Barcelona), Spain}

\begin{abstract}

We formulate higher derivative gravity with Lagrange multiplier
constraint and scalar projectors. Its gauge-fixed formulation as well as
vector fields formulation is developed and corresponding spontaneous Lorentz
symmetry breaking is investigated. We show that the only propagating mode is
higher derivative graviton while scalar and vector modes do not
propagate. Despite to higher derivatives structure of the action,
its first FRW equation is the first order differential equation which admits
the inflationary universe solution.

\end{abstract}


\maketitle

\section{Introduction \label{I}}

The study of the evolution of early-time and late-time universe indicates
that General Relativity is not complete theory of gravitational interaction.
At best, it may be the effective theory to describe the classical gravity
at intermediate energies. Indeed, there are quite strong proposals on the
unified description of the early-time inflation and late-time acceleration
in terms of modified gravity (for review, see \cite{Nojiri:2010wj}). It is
desirable to extend such modified gravity till Planck scale where it should
be renormalizable or somehow consistent (finite?) one at UV. Otherwise, such
proposals which pass local/cosmological tests (for recent review,
see \cite{CapozzielloFaraoni}) remain to be phenomenological ones.

It is quite well-known that higher derivative gravity may be multiplicatively
renormalizable (for a review, see \cite{Buchbinder:1992rb}).
However, the unitarity issue in such approach is so far the open problem.
Recently, very interesting attempt to formulation of power-counting
renormalizable gravity has been presented in terms of
gravity \cite{Horava:2009uw} which also uses effectively higher derivative
propagator. Unfortunately, such approach leads to explicit
breaking of Lorentz invariance from the very beginning.
Nevertheless, such formulation may be extended to covariant
theory \cite{Nojiri:2009th} where Lorentz invariance is broken spontaneously,
in the same way as celebrated gauge invariance.
The construction of higher-derivative power-counting model with consistent
graviton spectrum in this direction may suggest the new perspectives towards
to multiplicative renormalizability as well as resolution of some problems
which appear in Lorentz non-invariant gravities.

In the present work we propose covariant Lagrange multiplier constrained
higher derivative gravity with scalar projectors. Its gauge-fixed formulation
is developed. It is demonstrated that only higher derivative graviton degree
of freedom is propagating at tree level while scalar and vector degrees of
freedom do not propagate. The theory turns out to be
power-counting (super-)renormalizable. The spontaneous Lorentz symmetry
breaking is investigated. The equivalent representation in terms of
vectors is also developed.
Finally, FRW cosmology is studied. It is shown that theory looks very
similar to $R^2$ gravity. The possibility of eternal inflation is
demonstrated.

\section{The model of power-counting renormalizable,
covariant higher derivative gravity \label{SecII}}

It is expected that power-counting renormalizable covariant gravity should
be higher derivative theory, for instance, of the sort recently proposed in
refs.~\cite{Nojiri:2009th}.
In the present section we propose new formulation of such theory and derive
its gauge-fixed formulation and propagator. It will be shown that it may
have very good UV behavior in gauge-fixed formulation. Different versions
of Lagrange multiplier gravity were discussed in refs.~\cite{Nojiri:2010wj}.

In general, the unitarity is broken in the higher derivative theories.
In order to guarantee the unitarity, the models, where Lorentz symmetry and/or
the full general covariance
is explicitly broken, were proposed by Ho\v{r}ava \cite{Horava:2009uw}.
Although the models are very interesting, due to the lacking of the Lorentz symmetry
and/or the full general covariance, there appears an extra propagating
and problematic scalar mode (see \cite{extrascalar}, for example).
In the models proposed below, there is the Lorentz symmetry and/or
the full general covariance in the actions but the symmetry and/or the covariance
is broken spontaneously. As a result, we obtain the models where the
UV behavior of the graviton propagator is improved but any extra mode,
like scalar mode, does not appear.

We now start with the action including the Lagrange multiplier field
$\lambda$ \cite{vikman} and the scalar field $\phi$:
\be
\label{Lag}
S_\mathrm{Lag} = - \int d^4 x \sqrt{-g} \lambda \left( \frac{1}{2}
\partial_\mu \phi \partial^\mu \phi
+ U_0 \right) \, ,
\ee
which gives a constraint
\be
\label{LagHL2}
\frac{1}{2} \partial_\mu \phi \partial^\mu \phi
+ U_0 = 0\, ,
\ee
that is, the vector $(\partial_\mu \phi)$ is time-like.
Therefore the Lorentz symmetry and/or
the full general covariance is broken spontaneously.
The detailed discussion about the spontaneous breakdown of the symmetry and/or the
covariance will be discussed in detail in the next section.
For the spontaneous breakdown, $U_0$ needs not to be a constant
but some non-vanishing and positive function of $\phi$.  Just for the simplicity,
we consider the case that $U_0$ is a constant.
At least locally, one can choose the direction of time to be parallel to
$(\partial_\mu \phi)$.
Then Eq.~(\ref{LagHL2}) has the following form:
\be
\label{LagHL3}
\frac{1}{2} \left(\frac{d\phi}{dt}\right)^2 = U_0 \, .
\ee
Therefore the spatial region becomes a hypersurface where $\phi$ is a
constant since
the hypersurface is orthogonal to the vector $(\partial_\mu \phi)$.

We are interesting in the fluctuations over the flat background:
\be
\label{fluc}
g_{\mu\nu} = \eta_{\mu\nu} + h_{\mu\nu} \, .
\ee
Note that (\ref{LagHL3}) gives
\be
\label{ppert1}
\phi = \sqrt{2 U_0} t\, .
\ee
Then one can define a projection operator
\be
\label{ppert2}
P_\mu^{\ \nu} \equiv \delta_\mu^{\ \nu} + \frac{\partial_\mu \phi
\partial^\nu \phi}{2U_0}\, ,
\ee
and it follows
\bea
\label{ppert3}
P_i^{\ \mu} P_j^{\ \nu} R_{\mu\nu} &=&
\frac{1}{2}\left(h^\rho_{\ i,j\rho} + h^\rho_{\ j,i\rho} - \partial_\rho
\partial^\rho h_{ij,} - \partial_i \partial_j
\left( h_\rho^{\ \rho} \right) \right) \, , \\
\label{ppert4}
\frac{1}{2 U_0 } P_i^{\ \mu} P_j^{\ \nu}
\partial_\rho \phi \nabla^\rho \nabla_\mu \nabla_\nu \phi &=& - \frac{1}{2}
\left(h_{ti,jt} + h_{tj,it} - h_{ij,tt} \right) \, , \\
\partial^\mu \phi \partial^\nu \phi \nabla_\mu \nabla_\nu
+ 2 U_0 \nabla^\rho \nabla_\rho &=& 2 U_0 \partial_k \partial^k \, ,
\eea
and therefore
\be
\label{ppert5}
P_i^{\ \mu} P_j^{\ \nu} \left( R_{\mu\nu} - \frac{1}{2 U_0}
\partial_\rho \phi \nabla^\rho \nabla_\mu \nabla_\nu \phi
\right) =
\frac{1}{2}\left(h_{ki,j}^{\ \ \ \ k}
+ h_{kj,i}^{\ \ \ \ k} - h_{ij,k}^{\ \ \ \ k} - \partial_i \partial_j
\left( h_\mu^{\ \mu} \right) \right) \, . \\
\ee
Note that $P_0^{\ \mu} = 0$.
Then we can propose the action of the power-counting renormalizable,
covariant higher derivative gravity with scalar projector as\footnote{
We have merely found the expressions of the actions (\ref{ppert6})
and (\ref{ppert7}) so that
the propagator of the graviton could be given in (\ref{prpgtr}).
We have not found any deep physical  principle  to choose
the actions in the forms of (\ref{ppert6}) and (\ref{ppert7}) but
we believe that there might exist some hidden  symmetry to construct such theories.
}
\bea
\label{ppert6}
S_{2n+2} &=& \int d^4 x \sqrt{-g} \left[ \frac{R}{2\kappa^2} - \alpha
\left\{
\left(\partial^\mu \phi \partial^\nu \phi \nabla_\mu
\nabla_\nu - \partial_\mu \phi \partial^\mu \phi
\nabla^\rho \nabla_\rho \right)^n
P_\alpha^{\ \mu} P_\beta^{\ \nu} \left( R_{\mu\nu} - \frac{1}{2 U_0 }
\partial_\rho \phi \nabla^\rho \nabla_\mu \nabla_\nu \phi
\right) \right\} \right. \nn
&& \times \left\{
\left(\partial^\mu \phi \partial^\nu \phi \nabla_\mu
\nabla_\nu - \partial_\mu \phi \partial^\mu \phi
\nabla^\rho \nabla_\rho \right)^n
P^{\alpha\mu} P^{\beta\nu} \left( R_{\mu\nu} - \frac{1}{2 U_0 }
\partial_\rho \phi \nabla^\rho \nabla_\mu \nabla_\nu \phi
\right) \right\} \nn
&& \left. - \lambda \left( \frac{1}{2} \partial_\mu \phi \partial^\mu \phi
+ U_0 \right) \right]\, ,
\eea
for $z=2n + 2$ model $\left( n=0,1,2,\cdots\right)$,\footnote{
The idea proposed in ref.\cite{Horava:2009uw} for quantum gravity
is to modify the ultraviolet
behavior of the graviton propagator in Lorentz non-invariant way as
$1/\left|\bm{k}\right|^{2z}$, where $\bm{k}$ is the spatial momenta
and $z$ could be 2, 3 or larger integers.
They are defined  by the scaling properties of space-time
coordinates $\left(\bm{x},t\right)$ as $\bm{x}\to b\bm{x}$ and
$t\to b^z t$. When $z=3$, the theory seems to be UV renormalizable.
} and
\bea
\label{ppert7}
S_{2n+3} &=& \int d^4 x \sqrt{-g} \left[ \frac{R}{2\kappa^2} - \alpha
\left\{
\left(\partial^\mu \phi \partial^\nu \phi \nabla_\mu
\nabla_\nu - \partial_\mu \phi \partial^\mu \phi
\nabla^\rho \nabla_\rho \right)^n
P_\alpha^{\ \mu} P_\beta^{\ \nu} \left( R_{\mu\nu} - \frac{1}{2 U_0 }
\partial_\rho \phi \nabla^\rho \nabla_\mu \nabla_\nu \phi
\right) \right\} \right. \nn
&& \times \left\{
\left(\partial^\mu \phi \partial^\nu \phi \nabla_\mu
\nabla_\nu - \partial_\mu \phi \partial^\mu \phi
\nabla^\rho \nabla_\rho \right)^{n+1}
P^{\alpha\mu} P^{\beta\nu} \left( R_{\mu\nu} - \frac{1}{2 U_0 }
\partial_\rho \phi \nabla^\rho \nabla_\mu \nabla_\nu \phi
\right) \right\} \nn
&& \left. - \lambda \left( \frac{1}{2} \partial_\mu \phi \partial^\mu \phi
+ U_0 \right) \right]\, .
\eea
for $z=2n + 3$ model $\left( n=0,1,2,\cdots\right)$.
Here the quantity $z$ is introduced to express the anisotropy between the time
coordinate and spacial coordinates in (\cite{Horava:2009uw}).

One can confirm that the actions admit a flat space vacuum solution. Indeed,
field equations are:
\be
\label{sym4}
0 = \frac{1}{2\kappa^2} \left( R_{\mu\nu} - \frac{1}{2} g_{\mu\nu} R \right)
+ G^\mathrm{higher}_{\mu\nu} - \frac{\lambda}{2}
\partial_\mu \phi \partial_\nu \phi + \frac{1}{2}
g_{\mu\nu}
\left( \frac{1}{2} \partial_\rho \phi \partial^\rho \phi + U_0 \right)\, .
\ee
Here $G^\mathrm{higher}_{\mu\nu}$ comes from the higher derivative term (the
second term) in the actions (\ref{ppert6}) and (\ref{ppert7}) .
Assuming the flat vacuum solution, Eq.~(\ref{LagHL2}) given by the
variation over $\lambda$ has a form (\ref{LagHL3}).
Since for the flat space solution all the curvatures and $\nabla_\mu
\nabla_\nu \phi$ vanish, Eq.~(\ref{sym4}) reduces to
\be
\label{sym5}
0 = \lambda \partial_\mu \phi \partial_\nu \phi\, ,
\ee
whose solution is $\lambda=0$ since $\partial_\mu \phi\neq 0$ due to the
constraint
equation (\ref{LagHL2}) (if the coordinate system is chosen properly, we
have $\partial_t \phi = \sqrt{2U_0}$ and $\partial_i \phi = 0$).
Hence, the actions (\ref{ppert6}) and (\ref{ppert7}) admit the flat space
vacuum solution with $\lambda=0$.

Let us investigate the perturbation from the flat background (\ref{fluc})
with $\lambda=0$ in more detail.

First, by using the diffeomorphism invariance with respect to time
coordinate, we choose Eq.~(\ref{ppert1}) as a (unitary) gauge condition.
Then the actions (\ref{ppert6}) and (\ref{ppert7}) have the following form:
\bea
\label{pert2}
S_{2n+2} &=& \int d^4 x \left[ - \frac{1}{8\kappa^2} \left\{ - 2 h_{tt}
\left( \delta^{ij} \partial_k \partial^k - \partial^i \partial^j
\right) h_{ij}
+ 2 h_{ti} \left( \delta^{ij} \partial_k
\partial^k - \partial^i \partial^j \right) h_{tj}
\right. \right. \nn
&& + h_{ti} \left( 2 \delta^{jk} \partial^i - \delta^{ik}
\partial^j - \delta^{ij} \partial^k \right) \partial_t h_{jk}
+ h_{ij} \left( \left( \delta^{ij}
\delta^{kl} - \frac{1}{2} \delta^{ik} \delta^{jl} - \frac{1}{2}
\delta^{il} \delta^{jk} \right)
\left( - \partial_t^2 + \partial_k \partial^k\right) \right. \nn
&& \left.\left. - \delta^{ij} \partial^k \partial^l - \delta^{kl}
\partial^i \partial^j
+ \frac{1}{2} \left( \delta^{ik} \partial^j \partial^l + \delta^{il}
\partial^j \partial^k
+ \delta^{jk} \partial^i \partial^l + \delta^{jl} \partial^i \partial^k
\right) \right) h_{kl} \right\} \nn
&& - 2^{2n-2} \alpha U_0^{2n} \left\{ \left( \partial_k \partial^k \right)^n
\left(h_{ki,j}^{\ \ \ \ k}
+ h_{kj,i}^{\ \ \ \ k} - h_{ij,k}^{\ \ \ \ k} - \partial_i \partial_j
\left( h_\mu^{\ \mu} \right) \right) \right\} \nn
&& \left. \times \left\{ \left( \partial_k \partial^k \right)^n
\left(h_{k\ ,}^{\ i\ jk}
+ h_{k\ ,}^{\ j\ ik} - h_{\ \ ,k}^{ij\ \ k} - \partial^i \partial^j
\left( h_\mu^{\ \mu} \right) \right) \right\}
+ U_0 \lambda h_{tt} \right] \, , \\
\label{pert3}
S_{2n+3} &=& \int d^4 x \left[ - \frac{1}{8\kappa^2}
\left\{ - 2 h_{tt} \left( \delta^{ij} \partial_k
\partial^k - \partial^i \partial^j \right) h_{ij}
+ 2 h_{ti} \left( \delta^{ij} \partial_k \partial^k - \partial^i \partial^j
\right) h_{tj} \right. \right. \nn
&& + h_{ti} \left(
2 \delta^{jk} \partial^i - \delta^{ik} \partial^j - \delta^{ij}
\partial^k \right) \partial_t h_{jk}
+ h_{ij} \left( \left( \delta^{ij} \delta^{kl} - \frac{1}{2}
\delta^{ik} \delta^{jl} - \frac{1}{2} \delta^{il} \delta^{jk}
\right)
\left( - \partial_t^2 + \partial_k \partial^k\right) \right. \nn
&& \left.\left. - \delta^{ij} \partial^k \partial^l - \delta^{kl}
\partial^i \partial^j
+ \frac{1}{2} \left( \delta^{ik} \partial^j \partial^l + \delta^{il}
\partial^j \partial^k
+ \delta^{jk} \partial^i \partial^l + \delta^{jl} \partial^i \partial^k
\right) \right) h_{kl} \right\} \nn
&& - 2^{2n-1} \alpha U_0^{2n+1} \left\{ \left( \partial_k \partial^k
\right)^n
\left(h_{ki,j}^{\ \ \ \ k}
+ h_{kj,i}^{\ \ \ \ k} - h_{ij,k}^{\ \ \ \ k} - \partial_i \partial_j
\left( h_\mu^{\ \mu} \right) \right) \right\} \nn
&& \left. \times \left\{ \left( \partial_k \partial^k \right)^{n+1}
\left(h_{k\ ,}^{\ i\ jk}
+ h_{k\ ,}^{\ j\ ik} - h_{\ \ ,k}^{ij\ \ k} - \partial^i \partial^j
\left( h_\mu^{\ \mu} \right) \right) \right\}
+ U_0 \lambda h_{tt} \right] \, .
\eea
Here, only the terms quadratic with respect to the perturbation are kept.
Note that there remains the diffeomorphism invariance with respect to the
spatial coordinates.
It will be fixed later in (\ref{pert6}).
An important thing is that there does not appear $h_{ti}$ in the higher
derivative term with a coefficient $\alpha$.
The constraint equation (\ref{LagHL2}) shows
\be
\label{pert4}
h_{tt} = 0\, .
\ee
The variation of $h_{tt}$ can be solved with respect to $\lambda$:
\be
\label{pert5}
\lambda = - \frac{1}{4\kappa^2 U_0 } \left( \delta^{ij} \partial_k
\partial^k - \partial^i \partial^j \right) h_{ij}
+ 2^{2n-1} \alpha U_0^{2n-1} \left( \partial_k \partial^k \right)^{2n}
\partial^i \partial^j
\left(h_{ki,j}^{\ \ \ \ k}
+ h_{kj,i}^{\ \ \ \ k} - h_{ij,k}^{\ \ \ \ k} - \partial_i \partial_j
\left( h_\mu^{\ \mu} \right) \right) \, ,
\ee
for the action (\ref{ppert6}) and
\be
\label{pert5b}
\lambda = - \frac{1}{4\kappa^2 U_0 } \left( \delta^{ij} \partial_k
\partial^k - \partial^i \partial^j \right) h_{ij}
+ 2^{2n} \alpha U_0^{2n} \left( \partial_k \partial^k \right)^{2n+1}
\partial^i \partial^j \left(h_{ki,j}^{\ \ \ \ k}
+ h_{kj,i}^{\ \ \ \ k} - h_{ij,k}^{\ \ \ \ k} - \partial_i \partial_j
\left( h_\mu^{\ \mu} \right) \right) \, .
\ee
for the action (\ref{ppert7}).
The linearized equations given by the variation over $\phi$ are:
\be
\label{pert5c}
0 = \partial_t \left\{\lambda + 2^{2n - 1 } \alpha U_0^{2n - 1}
\left( \partial_k \partial^k \right)^{2n} \partial^i \partial^j
\left(h_{ki,j}^{\ \ \ \ k}
+ h_{kj,i}^{\ \ \ \ k} - h_{ij,k}^{\ \ \ \ k} - \partial_i \partial_j
\left( h_\mu^{\ \mu} \right) \right) \right\} \, ,
\ee
for the action (\ref{ppert6}) and
\be
\label{pert5d}
0 = \partial_t \left\{ \lambda + 2^{2n} \alpha U_0^{2n}
\left( \partial_k \partial^k \right)^{2n+1} \partial^i \partial^j
\left(h_{ki,j}^{\ \ \ \ k}
+ h_{kj,i}^{\ \ \ \ k} - h_{ij,k}^{\ \ \ \ k} - \partial_i \partial_j
\left( h_\mu^{\ \mu} \right) \right) \right\} \, .
\ee
for the action (\ref{ppert7}).

We now decompose $h_{ti}$, which corresponds to the fluctuation of the shift
function $N_i$, as follows
\be
\label{ppp1}
h_{ti}= \partial_i s + v_i\, ,\quad \partial^i v_i = 0\, .
\ee
Here $s$ is the spatial scalar.
We further write the linearized diffeomorphism invariance transformations
with respect to the spatial coordinates as follows
\be
\label{ppp2}
\delta x^i = \partial^i u + w^i\, ,\quad \partial_i w^i = 0 \, .
\ee
Then under the diffeomorphism, $s$ and $v_i$ in (\ref{ppp1}) are transformed
as
\be
\label{ppp3}
\delta s = \partial_t u\, ,\quad \delta v_i = \partial_t w_i \, ,
\ee
and therefore one can choose the gauge condition $s=v^i=0$, that is,
\be
\label{pert6}
h_{ti}=0\, .
\ee
Furthermore the variation of $h_{ti}$ gives
\be
\label{pert6b}
\partial_t \left( -2 \delta^{jk} \partial^i + \delta^{ik} \partial^j
+ \delta^{ij} \partial^k \right) h_{jk} = 0\, ,
\ee
which is identical with that in the usual Einstein gravity since the higher
derivative terms in the
actions (\ref{pert2}) and (\ref{pert3}) do not contain $h_{ti}$.
We decompose $h_{ij}$ as
\be
\label{pert7}
h_{ij} = \delta_{ij} A + \partial_j B_i + \partial_i B_j + C_{ij} + \left(
\partial_i \partial_j - \frac{1}{3}\delta_{ij}
\partial_k \partial^k \right) E\, ,
\ee
with
\be
\label{pert8}
\partial^i B_i = 0\, ,\quad \partial^i C_{ij} = \partial^j C_{ij} = 0
\, ,\quad C_i^{\ i}=0\, ,
\ee
Then substituting (\ref{pert7}) into (\ref{pert6b}), we obtain
\be
\label{pert8b}
0 = \partial_t \left( -4 \partial_i A + 2 \partial_k \partial^k B_i
+ \frac{4}{3}\partial_i \partial_k \partial^k E \right) \, .
\ee
By multiplying with $\partial^i$, one gets
\be
\label{ppp4}
\partial_t \partial_i \partial^i \left( - 4 A
+ \frac{4}{3} \partial_k \partial^k E \right) =0\, ,
\ee
which shows
\be
\label{ppp4b}
A = \frac{1}{3} \partial_k \partial^k E\, ,
\ee
under the boundary condition that $A$ and $E$ should vanish at spatial
infinity.
Then the equation given by substituting (\ref{ppp4b}) into (\ref{pert8b})
gives
\be
\label{ppp5}
\partial_t \partial_j \partial^j B_i = 0\, ,
\ee
which also indicates that $B_i=0$ under the boundary condition that $B_i$
should vanish at spatial infinity.

Using (\ref{pert7}) with (\ref{pert8}), Eqs.~(\ref{pert5})
and (\ref{pert5c}) have the following forms:
\bea
\label{ppp6}
\lambda &=& \frac{1}{2\kappa^2 U_0} \partial_k \partial^k \left( - A
+ \frac{1}{3} \partial_j \partial^j E \right) - 2^{2n} \alpha U_0^{2n-1}
\left( \partial_k \partial^k \right)^{2n + 2}
\left( - A + \frac{1}{3} \partial_j \partial^j E \right)\, , \\
\label{ppp7}
0 &=& \partial_t \left\{ \lambda + 2^{2n} \alpha U_0^{2n - 1}
\left( \partial_k \partial^k \right)^{2n+2} \left( - A + \frac{1}{3}
\partial_j \partial^j E \right) \right\}\, ,
\eea
and (\ref{pert5b}) and (\ref{pert5d})
\bea
\label{ppp8}
\lambda &=& \frac{1}{2\kappa^2 U_0} \partial_k \partial^k \left( - A
+ \frac{1}{3} \partial_j \partial^j E \right) - 2^{2n+1} \alpha U_0^{2n}
\left( \partial_k \partial^k \right)^{2n + 3}
\left( - A + \frac{1}{3} \partial_j \partial^j E \right)\, , \\
\label{ppp9}
0 &=& \partial_t \left\{ \lambda + 2^{2n+1} \alpha U_0^{2n}
\left( \partial_k \partial^k \right)^{2n+3} \left( - A + \frac{1}{3}
\partial_j \partial^j E \right) \right\}\, .
\eea
Combining (\ref{ppp4b}) with the above equations, we find
\be
\label{pert9}
\lambda = B_i = 0 \, ,
\ee
and therefore the scalar modes $\lambda$
and the vector mode $B_i$ do not propagate.

After the gauge fixing and using (\ref{pert9}) etc., the actions
(\ref{ppert6}) and (\ref{ppert7}) have the following forms:\footnote{
It is interesting that the structure of the actions (\ref{pert10o})
and (\ref{pert11o}) reminds the one of
$U(1)$ invariant $F(R)$ Ho\v{r}ava-Lifshitz gravity \cite{Kluson:2010za}. }
\bea
\label{pert10o}
S_{2n+2} &=& \int d^4 x \left[ \frac{1}{8\kappa^2} \left\{
C_{ij} \left( - \partial_t^2 + \partial_k \partial^k\right) C^{ij}
\right\} - 2^{2n-2} \alpha U_0^{2n} \left\{\left( \partial_k \partial^k
\right)^{n+1}C_{ij}\right\}
\left\{ \left( \partial_k \partial^k \right)^{n+1} C^{ij} \right\} \right. \nn
&& + \frac{1}{8\kappa^2} \left\{ - 6A \left( - \partial_t^2 + \partial_k
\partial^k\right) A
+ \frac{2}{3}\partial_k \partial^k E \left( - \partial_t^2 + \partial_k
\partial^k\right)
\partial_k \partial^k E \right. \nn
&& \left. + 4 A \partial_k \partial^k A
+ \frac{4}{3} A \left( \partial_k \partial^k \right)^2 E - \frac{8}{9}
\partial_k \partial^k E \left( \partial_k \partial^k \right)^2 E
\right\} \nn
&& - 2^{2n-2} \alpha U_0^{2n} \left\{\left( \partial_k \partial^k
\right)^n \left(- \partial_i \partial_j A - \delta_{ij} \partial_k \partial^k A
+ \frac{1}{3}\partial_i \partial_j \partial_k \partial^k E
+ \frac{1}{3} \delta_{ij} \left( \partial_k \partial^k \right)^2 E \right)
\right\} \nn
&& \left. \times \left\{\left( \partial_k \partial^k
\right)^n \left(- \partial^i \partial^j A - \delta^{ij} \partial_k \partial^k A
+ \frac{1}{3}\partial^i \partial^j \partial_k \partial^k E
+ \frac{1}{3} \delta^{ij} \left( \partial_k \partial^k \right)^2 E \right)
\right\} \right] \, , \\
\label{pert11o}
S_{2n+3} &=& \int d^4 x \left[ \frac{1}{8\kappa^2} \left\{
C_{ij} \left( - \partial_t^2 + \partial_k \partial^k\right)
C^{ij} \right\} - 2^{2n-1} \alpha U_0^{2n+1} \left\{ \left(
\partial_k \partial^k \right)^{n+1}C_{ij} \right\}
\left\{ \left( \partial_k \partial^k \right)^{n+2} C^{ij}\right\} \right. \nn
&& + \frac{1}{8\kappa^2} \left\{ - 6A \left( - \partial_t^2 + \partial_k
\partial^k\right) A
+ \frac{2}{3}\partial_k \partial^k E \left( - \partial_t^2 + \partial_k
\partial^k\right)
\partial_k \partial^k E \right. \nn
&& \left. + 4 A \partial_k \partial^k A
+ \frac{4}{3} A \left( \partial_k \partial^k \right)^2 E - \frac{8}{9}
\partial_k \partial^k E \left( \partial_k \partial^k \right)^2 E
\right\} \nn
&& - 2^{2n-1} \alpha U_0^{2n+1} \left\{\left( \partial_k \partial^k
\right)^n \left(- \partial_i \partial_j A - \delta_{ij} \partial_k \partial^k A
+ \frac{1}{3}\partial_i \partial_j \partial_k \partial^k E
+ \frac{1}{3} \delta_{ij} \left( \partial_k \partial^k \right)^2 E \right)
\right\} \nn
&& \left. \times \left\{\left( \partial_k \partial^k
\right)^{n+1} \left(- \partial^i \partial^j A - \delta^{ij} \partial_k
\partial^k A
+ \frac{1}{3}\partial^i \partial^j \partial_k \partial^k E
+ \frac{1}{3} \delta^{ij} \left( \partial_k \partial^k \right)^2 E \right)
\right\} \right] \, .
\eea
Then by the variation of $A$, we obtain
\bea
\label{AAA1}
&& 0 = + \frac{1}{8\kappa^2} \left\{ - 12 \left( - \partial_t^2 + \partial_k
\partial^k\right) A
+ 8 \partial_k \partial^k A + \frac{4}{3} \left( \partial_k \partial^k
\right)^2 E \right\} \nn
&& - 2^{2n-1} \alpha U_0^{2n}
\left(- \partial_i \partial_j - \delta_{ij} \partial_k \partial^k \right)
\left\{\left( \partial_k \partial^k\right)^{2n}
\left(- \partial^i \partial^j A - \delta^{ij} \partial_k \partial^k A
+ \frac{1}{3}\partial^i \partial^j \partial_k \partial^k E
+ \frac{1}{3} \delta^{ij} \left( \partial_k \partial^k \right)^2 E \right)
\right\} \, ,
\eea
for the action (\ref{pert10o}) and
\bea
\label{AAA2}
&& 0 = + \frac{1}{8\kappa^2} \left\{ - 12 \left( - \partial_t^2 + \partial_k
\partial^k\right) A
+ 8 \partial_k \partial^k A + \frac{4}{3} \left( \partial_k \partial^k
\right)^2 E \right\} \nn
&& - 2^{2n} \alpha U_0^{2n+1}
\left(- \partial_i \partial_j - \delta_{ij} \partial_k \partial^k \right)
\left\{\left( \partial_k \partial^k\right)^{2n+1} \left(- \partial^i \partial^j
A - \delta^{ij} \partial_k \partial^k A
+ \frac{1}{3}\partial^i \partial^j \partial_k \partial^k E
+ \frac{1}{3} \delta^{ij} \left( \partial_k \partial^k \right)^2 E \right)
\right\} \, ,
\eea
for the action (\ref{pert11o}).
On the other hand, by the variation over $E$, one gets
\bea
\label{EEE1}
&& 0 = \partial_k \partial^k \left[ \frac{1}{8\kappa^2}
\left\{ \frac{4}{3}\left( - \partial_t^2 + \partial_k \partial^k\right)
\partial_k \partial^k E
+ \frac{4}{3} \partial_k \partial^k A
+ \frac{16}{9} \left( \partial_k \partial^k \right)^2 E \right\} \right. \nn
&& \left. + \frac{2^{2n-1}}{3} \alpha U_0^{2n}
\left(- \partial_i \partial_j - \delta_{ij} \partial_k \partial^k \right)
\left\{\left( \partial_k \partial^k\right)^{2n}
\left(- \partial^i \partial^j A - \delta^{ij} \partial_k \partial^k A
+ \frac{1}{3}\partial^i \partial^j \partial_k \partial^k E
+ \frac{1}{3} \delta^{ij} \left( \partial_k \partial^k \right)^2 E \right)
\right\} \right] \, ,
\eea
for the action (\ref{pert10o}) and
\bea
\label{EEE2}
&& 0 = \partial_k \partial^k \left[ \frac{1}{8\kappa^2}
\left\{ \frac{4}{3}\left( - \partial_t^2 + \partial_k \partial^k\right)
\partial_k \partial^k E
+ \frac{4}{3} \partial_k \partial^k A
+ \frac{16}{9} \left( \partial_k \partial^k \right)^2 E \right\} \right. \nn
&& \left. + \frac{2^{2n}}{3} \alpha U_0^{2n+1}
\left(- \partial_i \partial_j - \delta_{ij} \partial_k \partial^k \right)
\left\{\left( \partial_k \partial^k\right)^{2n+1} \left(- \partial^i \partial^j
A - \delta^{ij} \partial_k \partial^k A
+ \frac{1}{3}\partial^i \partial^j \partial_k \partial^k E
+ \frac{1}{3} \delta^{ij} \left( \partial_k \partial^k \right)^2 E \right)
\right\} \right] \, ,
\eea
for the action (\ref{pert11o}).
By using (\ref{ppp4b}), both of Eqs.~(\ref{AAA1}), (\ref{AAA2}), (\ref{EEE1}),
and (\ref{EEE2}) give the same expression:
\be
\label{AAA3}
0 = \partial_t^2 A \, .
\ee
Therefore $A$ and $E$ only depend on the spacial coordinate and therefore they
do not propagate.
Then we have shown that all the scalar modes $\phi$,
$\lambda$, $h_{tt}$, $s$, $A$, and $E$ and all the vector modes $v_i$ and
$B_i$ in Eqs.~(\ref{ppp1}) and (\ref{pert7}) do not propagate and
the only propagating mode is massless graviton corresponding to
the tensor mode $C_{ij}$, which should be distinguished from
Ho\v{r}ava quantum gravity \cite{Horava:2009uw} where Lorentz invariance is
explicitly broken.
Then the actions (\ref{ppert6}) and (\ref{ppert7}) and therefore
(\ref{pert10o}) and (\ref{pert11o}) reduce to the simple forms:
\bea
\label{pert10}
S_{2n+2} &=& \int d^4 x \left[ \frac{1}{8\kappa^2} \left\{
C_{ij} \left( - \partial_t^2 + \partial_k \partial^k\right) C^{ij}
\right\} - 2^{2n-2} \alpha U_0^{2n} \left\{\left( \partial_k \partial^k
\right)^{n+1}C_{ij}\right\}
\left\{ \left( \partial_k \partial^k \right)^{n+1} C^{ij} \right\} \right]
\, , \\
\label{pert11}
S_{2n+3} &=& \int d^4 x \left[ \frac{1}{8\kappa^2} \left\{
C_{ij} \left( - \partial_t^2 + \partial_k \partial^k\right) C^{ij}
\right\} - 2^{2n-1} \alpha U_0^{2n+1} \left\{ \left( \partial_k \partial^k
\right)^{n+1}C_{ij} \right\}
\left\{ \left( \partial_k \partial^k \right)^{n+2} C^{ij}\right\} \right]
\, .
\eea
In the original models \cite{Nojiri:2009th}, due to the
traceless and transverse conditions for $C_{ij}$ in (\ref{pert8}),
the higher order terms do not contribute to the propagator
of the graviton and therefore the UV behavior has not been really improved.

Thus, the propagator has the following form in the momentum space:
\bea
\label{prpgtr}
&& \left< h_{ij}(p) h_{kl}(-p) \right>
= \left< C_{ij}(p) C_{kl}(-p) \right> \nn
&& = \frac{1}{2} \left\{ \left( \delta_{ij} - \frac{p_i
p_j}{\bm{p}^2}\right)
\left( \delta_{kl} - \frac{p_k p_l}{\bm{p}^2}
\right) - \left( \delta_{ik} - \frac{p_i p_k}{\bm{p}^2}\right)
\left( \delta_{jl} - \frac{p_j p_l}{\bm{p}^2}
\right) - \left( \delta_{il} - \frac{p_i p_l}{\bm{p}^2}\right)
\left( \delta_{jk} - \frac{p_j p_k}{\bm{p}^2}\right) \right\} \nn
&& \times \left\{
\begin{array}{ll}
\left( p^2 - 2^{2n} \alpha \kappa^2 U_0^{2n} \bm{p}^{4 ( n+1 )} \right)^{-1}
\, , & z=2n+2\ \mbox{case} \\
\left( p^2 - 2^{2n-1} \alpha \kappa^2 U_0^{2n+1} \bm{p}^{2 ( 2n+3 )}
\right)^{-1}
\, , & z=2n+3\ \mbox{case}
\end{array} \right. \, .
\eea
Here ${\bm{p}}^2 = \sum_{i=1}^3 \left( p^i \right)^2$ and
$p^2 = - \left(p^0\right)^2 + {\bm{p}}^2$.
If $\alpha>0$, there appears the tachyonic pole when
\be
\label{tachyon}
\begin{array}{ll}
1 = 2^{2n} \alpha \kappa^2 U_0^{2n} \bm{p}^{4 n + 2} \, , & z=2n+2\
\mbox{case} \\
1 = 2^{2n-1} \alpha \kappa^2 U_0^{2n+1} \bm{p}^{4 ( n+1 )} \, , & z=2n+3\
\mbox{case}
\end{array} \, ,
\ee
with $p^0=0$ and therefore at least the flat vacuum becomes unstable.
On the other hand, there exist a stable flat vacuum when $\alpha<0$.

In the present model, there is no propagating vector or scalar mode at least
on the tree level. The change of the tensor structure of the propagator in
(\ref{prpgtr}) means that the vector or scalar mode could appear, that is,
the vector or scalar mode must correspond to a composite state, which
usually does not appear at any perturbative level.
Therefore, it is expected the tensor structure should not be changed by the
quantum corrections.

In the ultraviolet region, where $\bm{k}$ is large,
the propagator behaves as $1/\left| \bm{k} \right|^4$ for $z=2$ ($n=0$) case
in (\ref{pert10}) and therefore the ultraviolet behavior is rendered.
For $z=3$ ($n=0$) case in (\ref{pert11}),
the propagator behaves as $1/\left| \bm{k} \right|^6$ and therefore
the model becomes power-counting renormalizable.
For $z=2n +2$ ($n\geq 1$) case in (\ref{pert10}) or $z=2n+3$ ($n\geq 1$)
case in (\ref{pert11}), the model becomes power-counting
super-renormalizable.
The dispersion relation of the graviton is then given by
\be
\label{sym29}
\omega = c_0 k^z\, ,
\ee
in the high energy region.
Here $c_0$ is a constant, $\omega$ is the angular frequency corresponding to
the energy and $k$ is the wave number corresponding to momentum.
If $c_0<0$, the dispersion relation becomes inconsistent and therefore
$c_0$ should be positive.

Let us discuss the generality of the expression (\ref{ppert5}), which
appears in the actions (\ref{ppert6}) and (\ref{ppert7}).
We now require for this expression, when a flat background is chosen
\begin{enumerate}
\item\label{1} The expression does not vanish. This condition is trivial but
this often occurs due to some identity and the condition without torsion
$\nabla_\mu g_{\nu\rho}$.
\item\label{2} The expression is given by the second rank symmetric tensor
as in (\ref{ppert5}).
\item\label{3} Each term contains the second derivative of perturbed metric
$h_{\mu\nu}$ like $\partial_\rho \partial_\sigma h_{\mu\nu}$.
This condition plays a role of the dimensionality
condition. Since $\partial_\mu \phi$ can be regarded as a constant vector,
the third derivative of $\phi$ like $\nabla_\mu \nabla_\nu \nabla_\rho \phi$
contains the second derivative of the perturbed metric like $\partial_\rho
\partial_\sigma h_{\mu\nu}$.
In order to obtain second rank symmetric tensor from the third rank tensor,
which is now the third derivative of $\phi$ like
$\nabla_\mu \nabla_\nu \nabla_\rho \phi$, we need one
more derivative of $\phi$ as in the second term of (\ref{ppert5}) like
$\nabla^\mu \phi \nabla_\mu \nabla_\nu \nabla_\rho \phi$,
$\nabla^\rho \left(\nabla_\mu \nabla_\nu \nabla_\rho \phi
+ \nabla_\nu \nabla_\mu \nabla_\rho \phi\right)$, or
$\nabla^\rho \left( \nabla_\mu \nabla_\rho \nabla_\nu \phi
+ \nabla_\nu \nabla_\rho \nabla_\mu \phi\right)$. This kind of terms was not
considered in \cite{Nojiri:2009th}.
\item\label{4} The expression does not contain $h_{ti}$ so that the
traceless and transverse conditions $C_i^{\ j}= \partial^i C_{ij} =0$
for $C_{ij}$ in (\ref{pert8}) should be kept as in the Einstein gravity.
This condition ensures the absence of the extra modes.
\item\label{5} The expression should contain the graviton, which is the
traceless and transverse part $C_{ij}$ in (\ref{pert7}) of the perturbed
metric.
\end{enumerate}
The combination of the terms in (\ref{ppert5}) satisfies the above
conditions but the combination in (\ref{ppert5}) is not, however, unique.
More generally, one can propose the combination like
\bea
\label{pppert1}
\mathcal{Q}_{\alpha\beta}
&\equiv& P_\alpha^{\ \mu} P_\beta^{\ \nu} \left(
R_{\mu\nu} - \frac{1}{2 U_0 } \partial_\rho
\phi \nabla^\rho \nabla_\mu \nabla_\nu \phi \right)
+ \left( c_1 \delta_\alpha^{\ \mu} \delta_\beta^{\ \nu}
+ c_2 P_\alpha^{\ \mu} P_\beta^{\ \nu} \right)
\nabla^\rho \left(\nabla_\mu \nabla_\nu \nabla_\rho \phi
+ \nabla_\nu \nabla_\mu \nabla_\rho \phi\right) \nn
&& + \left( c_3 \delta_\alpha^{\ \mu} \delta_\beta^{\ \nu}
+ c_4 P_\alpha^{\ \mu} P_\beta^{\ \nu} \right)
\nabla^\rho \left( \nabla_\mu \nabla_\rho \nabla_\nu \phi
+ \nabla_\nu \nabla_\rho \nabla_\mu \phi\right)
+ \left( c_5 g_{\mu\nu} + c_6 P_{\mu\nu} \right)
\left(\partial^\mu \phi \partial^\nu \phi R_{\mu\nu} + U_0 R\right)\, .
\eea
Here $c_1$, $c_2$, $c_3$, $c_4$, $c_5$, and $c_6$ are constants.
When we consider the fluctuation from the flat metric in (\ref{fluc}),
\bea
\label{pppert2}
&& \nabla^\rho \left(\nabla_\mu \nabla_\nu \nabla_\rho \phi
+ \nabla_\nu \nabla_\mu \nabla_\rho \phi\right)
\sim \nabla^\rho \left( \nabla_\mu \nabla_\rho \nabla_\nu \phi
+ \nabla_\nu \nabla_\rho \nabla_\mu \phi\right)
\sim - h_{tt,\mu\nu} \, , \nn
&& \partial^\mu \phi \partial^\nu \phi R_{\mu\nu} + U_0 R \sim
U_0 \left\{ \partial^i \partial^j h_{ij} - \partial_k \partial^k
\left(\delta^{ij} h_{ij} \right) \right\}\, .
\eea
Then by using (\ref{pert4}) and (\ref{pert7}) with (\ref{pert8})
and (\ref{pert9}), we find the terms with the coefficients
$c_1$, $c_2$, $c_3$, $c_4$, $c_5$, and $c_6$ vanish and therefore
these terms do not affect the propagator of graviton although they may
give rise some interaction terms. Having in mind the above considerations, one
may speculate that proposed theory is
multiplicatively-renormalizable in gauge-fixed formulation (after spontaneous
breaking of Lorentz symmetry).

In the above analysis, we have chosen Eq.~(\ref{ppert1}) as a gauge
condition.
Then Eq.~(\ref{pert4}) follows from the constraint equation (\ref{LagHL2}).
As another gauge condition, instead of (\ref{ppert1}), we may choose
(\ref{pert4}). When we write the fluctuation of $\phi$ as
\be
\label{pppert3}
\phi = \sqrt{2 U_0} t + \delta\phi\, ,
\ee
the constraint equation (\ref{LagHL2}) gives
\be
\label{pppert4}
\frac{\partial \delta \phi}{\partial t} = 0\, .
\ee
Then $\delta \phi$ does not depend of the time coordinate:
\be
\label{pppert5}
\delta \phi = \Phi(\bm{x})\, .
\ee
The gauge condition (\ref{pert4}) has the residual gauge symmetry for the
diffeomorphism with respect to the time coordinate $t$.
The residual gauge symmetry is the shift of $t$ by
a function independent of $t$
\be
\label{pppert6}
t \to t + f(\bm{x})\, .
\ee
Therefore by using the residual gauge transformation given by
\be
\label{pppert7}
f(\bm{x}) = - \frac{\Phi(\bm{x})}{\sqrt{2 U_0}}\, ,
\ee
we obtain (\ref{ppert1}) from (\ref{pppert3}).

Thus, we presented gauge-fixed formulation of the theory which leads to only
propagating higher derivative graviton. Scalar and vector degrees of freedom do
not propagate.

\section{Spontaneous Lorentz symmetry breaking}

Due to the constraint equation (\ref{LagHL2}), vector $\partial_\mu\phi$
has nontrivial value and therefore the Lorentz symmetry is broken. Note that
the Lorentz symmetry breaking is spontaneous.
Let us compare this situation with that in the usual field theory like the
Higgs model.
The usual $U(1)$ Higgs model, whose potential is given by
\be
\label{sym14}
V_\mathrm{Higgs} = - \frac{m^2}{2} \phi^* \phi
+ \frac{\lambda_0^2}{4} \left(\phi^* \phi \right)^2\, ,
\ee
has a global $U(1)$ symmetry, which is the invariance under the
transformation
\be
\label{sym15}
\phi \to \e^{i\theta_0} \phi\, ,
\ee
with a constant real parameter $\theta_0$.
In (\ref{sym14}), $\phi$ is a complex scalar field and $m$ and $\lambda_0$
are positive parameters.
The minimum of the potential is given by
\be
\label{sym16}
\phi = \frac{\e^{i\varphi}m}{\lambda_0}\, .
\ee
Here $\varphi$ is a constant phase. The value of $\varphi$ can be arbitrary.
If one chooses specific value of $\varphi$, the value of $\varphi$ is
changed under the $U(1)$ transformation (\ref{sym15}) as
$\varphi \to \varphi + \theta$,
and therefore the ground state is not invariant under the $U(1)$
transformation (\ref{sym15}) and the $U(1)$ symmetry breaks spontaneously.
One can always choose the real axis of the complex $\phi$-plane to be
parallel with the value of $\phi$ in the ground state so that $\varphi=0$.

In our model, the constraint equation (\ref{LagHL2}) shows that the value of
the vector $(\partial_\mu \phi)$ is located on the hyperboloid defined by
\be
\label{sym17}
 - x^\mu x_\mu \equiv t^2 - \bm{x}^2 = 2 U_0 \, .
\ee
The value of the vector $(\partial_\mu \phi)$ changes on the
hyperboloid under the Lorentz transformation. If we choose a value of
$(\partial_\mu \phi)$, the Lorentz symmetry is broken spontaneously.
After that one can always choose the time axis
to be parallel to the vector $(\partial_\mu \phi)$.

We should also note that the actions (\ref{ppert6}) and (\ref{ppert7}) have
a shift symmetry
\be
\label{sym3}
\phi \to \phi + \phi_0 \, .
\ee
Here $\phi_0$ is a constant.
In the flat vacuum background, the actions (\ref{ppert6}) and (\ref{ppert7})
are invariant under the time translation:
\be
\label{sym7}
t \to t + t_0\, .
\ee
Here $t_0$ is a constant.
Since the solution (\ref{ppert1}) of the constraint equation (\ref{LagHL3})
depends on the time coordinate $t$, the solution spontaneously
breaks the symmetry under the time translation (\ref{sym7}).
The solution (\ref{ppert1}) also breaks the shift symmetry in (\ref{sym3}).
Note that the diagonal symmetry of the time
translation (\ref{sym7}) and the shift symmetry (\ref{sym3}) is not broken.
In fact, if $\phi_0$ in (\ref{sym3}) is chosen as
\be
\label{sym8}
\phi_0 = - \sqrt{2U_0} t_0 \, ,
\ee
under the simultaneous transformation, the solution (\ref{ppert1}) is
invariant.
The diagonal symmetry effectively plays the role of the time translation and
the flat vacuum solution is effectively invariant under the time
translation.

\section{Vector field formulation}

Let us reformulate the theory in terms of vector field.
Note that the actions (\ref{ppert6}) and (\ref{ppert7}) can be rewritten by
using the vector field
$A_\mu$. For the vector filed $A_\mu$, if we impose the constraint
\be
\label{A1}
0= F_{\mu\nu} \equiv \partial_\mu A_\nu - \partial_\nu A_\mu\, ,
\ee
the vector field $A_\mu$ becomes pure gauge field and can be rewritten as
\be
\label{A2}
A_\mu = \partial_\mu \phi\, .
\ee
Furthermore, by introducing a new scalar field $\varphi$ and considering the
combination
\be
\label{A3}
A^\varphi_\mu \equiv A_\mu + \partial_\mu \varphi\, ,
\ee
the combination $A^\varphi_\mu$ is invariant under the gauge transformation
\be
\label{A4}
A_\mu \to A_\mu + \partial_\mu \epsilon (x^\mu)\, ,\quad
\varphi \to \varphi - \epsilon\, .
\ee
Then we rewrite the actions (\ref{ppert6}) and (\ref{ppert7}), which are
invariant under the gauge transformation (\ref{A4}) as
\bea
\label{Appert6}
S^A_{2n+2} &=& \int d^4 x \sqrt{-g} \left[ \frac{R}{2\kappa^2} - \alpha
\left\{
\left(A^{\varphi\, \mu} A^{\varphi\, \nu} \nabla_\mu \nabla_\nu - 2U_0
\nabla^\rho \nabla_\rho \right)^n
P_\alpha^{A\, \mu} P_\beta^{A\, \nu} \left( R_{\mu\nu} - \frac{1}{2 U_0 }
A^\varphi_\rho \nabla^\rho \nabla_\mu A^\varphi_\nu
\right) \right\} \right. \nn
&& \times \left\{
\left(A^{\varphi\,\mu} A^{\varphi\,\nu} \nabla_\mu \nabla_\nu - 2 U_0
\nabla^\rho \nabla_\rho \right)^n
P^{A\,\alpha\mu} P^{A\, \beta\nu} \left( R_{\mu\nu} - \frac{1}{2 U_0 }
A^\varphi_\rho \nabla^\rho \nabla_\mu A^\varphi_\nu
\right) \right\} \nn
&& \left. - \lambda \left( \frac{1}{2} A^\varphi_\mu A^{\varphi\,\mu}
+ U_0 \right) - B^{\mu\nu} F_{\mu\nu} \right]\, ,
\eea
for $z=2n + 2$ model $\left( n=0,1,2,\cdots\right)$, and
\bea
\label{Appert7}
S^A_{2n+3} &=& \int d^4 x \sqrt{-g} \left[ \frac{R}{2\kappa^2} - \alpha
\left\{
\left(A^{\varphi\, \mu} A^{\varphi\, \nu} \nabla_\mu \nabla_\nu - 2 U_0
\nabla^\rho \nabla_\rho \right)^n
P_\alpha^{\ \mu} P_\beta^{\ \nu} \left( R_{\mu\nu} - \frac{1}{2 U_0 }
A^\varphi_\rho \nabla^\rho \nabla_\mu A^\varphi_\nu
\right) \right\} \right. \nn
&& \times \left\{
\left(A^{\varphi\,\mu} A^{\varphi\,\nu} \nabla_\mu \nabla_\nu - 2 U_0
\nabla^\rho \nabla_\rho \right)^{n+1}
P^{\varphi\,\alpha\mu} P^{\varphi\,\beta\nu} \left( R_{\mu\nu} - \frac{1}{2 U_0 }
A^\varphi_\rho \nabla^\rho \nabla_\mu A^\varphi_\nu
\right) \right\} \nn
&& \left. - \lambda \left( \frac{1}{2} A^\varphi_\mu A^{\varphi\,\mu}
+ U_0 \right) - B^{\mu\nu} F_{\mu\nu} \right]\, .
\eea
for $z=2n + 3$ model $\left( n=0,1,2,\cdots\right)$.
Here $B^{\mu\nu}$ is the anti-symmetric tensor field whose variation gives a
constraint (\ref{A1}).
In the actions (\ref{Appert6}) and (\ref{Appert7}), $P^{A\, \nu}_\mu$ is
defined by
\be
\label{A5}
P^{A\, \nu}_\mu \equiv \delta_\mu^{\ \nu} + \frac{A^\varphi_\mu
A^{\varphi\,\nu}}{2U_0}\, ,
\ee
instead of (\ref{ppert2}).

The actions (\ref{Appert6}) and (\ref{Appert7}) are classically equivalent
to the actions (\ref{ppert6}) and (\ref{ppert7}). To be sure, let us check
the propagating modes on the flat background.
For this purpose, we decompose $B_{\mu\nu}$ and $A_\mu$ as follows,
\bea
\label{AA1}
&& B_{ti}= \partial_i f + e_i\, ,\quad B_{ij} = \partial_i m_j - \partial_j
m_i + \epsilon_{ijk}\partial^k k\, , \quad A_i = \partial_i + l_i\, ,\nn
&& \partial^i e_i = \partial^i m_i = \partial^i l_i\, .
\eea
Then one gets
\be
\label{AA2}
B^{\mu\nu} F_{\mu\nu}
= 2 f \partial_k \partial^k \left( - \partial_t \phi + A_t \right) - 2m^j
\partial_k \partial^k l_j + 2e^i \partial_t l_i
+ \mbox{total derivative terms} \, .
\ee
Note that in the above expression and therefore even in the total action,
there does not appear $e_i$ and $k$.
The variation over $f$ gives
\be
\label{AA3}
A_t = \partial_t \phi\, ,
\ee
and the variation over $m_i$ gives
\be
\label{AA4}
l_i = 0\, .
\ee
The variation over $e_i$ gives $\partial_t l_i = 0$, which is consistent
with (\ref{AA4}) and does not lead to new constraint.
On the other hand, the equation given by the variation over $A_0$ can be
solved with respect to $f$ and therefore $f$ becomes auxiliary field.
Similarly, the equation given by the variation over $l_i$ can be solved with
respect to $m_i$, which also becomes auxiliary field.
Then from (\ref{AA1}) and (\ref{AA3}), we find $A_\mu$ is pure
gauge field $A_\mu = \partial_\mu \phi$. Hence, in the actions (\ref{Appert6})
and (\ref{Appert7}), the scalar mode $\phi$ and the scalar field $\varphi$
appear only in the combination of $\partial_\mu \left( \phi + \varphi
\right)$.
Therefore the variations over $\phi$ and $\varphi$ give the identical
equations corresponding to (\ref{pert5c}) and (\ref{pert5d}). Hence, using
the same arguments as in Section \ref{SecII},
we obtain the same graviton propagator (\ref{prpgtr}).

\section{FRW cosmology}

In this section, we discuss simple accelerating FRW cosmology for
higher-derivative gravity under discussion.
We start with a little bit different but general action:
\bea
\label{cosppert6}
S &=& \int d^4 x \sqrt{-g} \left[
\frac{R}{2\kappa^2} - \sum_{n=0}^{n_\mathrm{max}} \alpha_n \left\{
\left(\partial^\mu \phi \partial^\nu \phi \nabla_\mu \nabla_\nu - \partial_\mu \phi
\partial^\mu \phi \nabla^\rho \nabla_\rho \right)^n
P_\alpha^{\ \mu} P_\beta^{\ \nu} \left( R_{\mu\nu} - \frac{1}{2 U_0 }
\partial_\rho \phi \nabla^\rho \nabla_\mu \nabla_\nu \phi
\right) \right\} \right. \nn
&& \times \left\{ \left(\partial^\mu \phi \partial^\nu \phi
\nabla_\mu \nabla_\nu - \partial_\mu \phi \partial^\mu \phi
\nabla^\rho \nabla_\rho \right)^n
P^{\alpha\mu} P^{\beta\nu} \left( R_{\mu\nu} - \frac{1}{2 U_0 }
\partial_\rho \phi \nabla^\rho \nabla_\mu \nabla_\nu \phi
\right) \right\} \nn
&& - \sum_{m=0}^{m_\mathrm{max}} \tilde\alpha \left\{
\left(\partial^\mu \phi \partial^\nu \phi \nabla_\mu \nabla_\nu - \partial_\mu \phi
\partial^\mu \phi \nabla^\rho \nabla_\rho \right)^m
P_\alpha^{\ \mu} P_\beta^{\ \nu} \left( R_{\mu\nu} - \frac{1}{2 U_0 }
\partial_\rho \phi \nabla^\rho \nabla_\mu \nabla_\nu \phi
\right) \right\} \nn
&& \times \left\{
\left(\partial^\mu \phi \partial^\nu \phi \nabla_\mu \nabla_\nu - \partial_\mu \phi
\partial^\mu \phi \nabla^\rho \nabla_\rho \right)^{m+1}
P^{\alpha\mu} P^{\beta\nu} \left( R_{\mu\nu} - \frac{1}{2 U_0 } \partial_\rho \phi
\nabla^\rho \nabla_\mu \nabla_\nu \phi
\right) \right\} \nn
&& \left. - \lambda \left( \frac{1}{2} \partial_\mu \phi \partial^\mu \phi
+ U_0 \right) \right]\, .
\eea
In the low energy, one only needs to consider the contribution from the
first Einstein-Hilbert term.
Since the $n=0$ term contributes in the next-to-leading order,
we now consider the simplified model given by the Einstein-Hilbert term and
the $n=0$ term:
\bea
\label{cos1}
S &=& \int d^4 x \sqrt{-g} \left[ \frac{R}{2\kappa^2} - \alpha_0
P_\alpha^{\ \mu} P_\beta^{\ \nu} \left( R_{\mu\nu} - \frac{1}{2 U_0 }
\partial_\rho \phi \nabla^\rho \nabla_\mu \nabla_\nu \phi \right)
P^{\alpha\mu} P^{\beta\nu} \left( R_{\mu\nu} - \frac{1}{2 U_0 }
\partial_\rho \phi \nabla^\rho \nabla_\mu \nabla_\nu \phi
\right) \right. \nn
&& \left. - \lambda \left( \frac{1}{2} \partial_\mu \phi \partial^\mu \phi
+ U_0 \right) \right]\, .
\eea
It is interesting that even in the FRW background, we can choose the local
Lorentz frame and obtain the same graviton propagator (\ref{prpgtr})
at short distances, where space-time can be regarded to be flat.

In order to consider the cosmology, we assume the FRW-like metric:
\be
\label{bFRW}
ds^2 = - \e^{2b(t)} dt^2 + a(t)^2 \sum_{i=1,2,3} \left(dx^i\right)^2\, .
\ee
Here $a(t)$ is called a scale factor.
The variation of the action with respect to $b(t)$ gives the equation
corresponding to the first FRW equation
and the variation with respect to $a(t)$ gives the equation corresponding to
the second FRW equation.
Since we have
\be
\label{gammas}
\Gamma^t_{tt} = \dot b \, ,\quad \Gamma^t_{ij} = \e^{-2b} a^2 H
\delta_{ij}\, ,\quad
\Gamma^i_{tj} = \Gamma^i_{jt} = H \delta^i_{\ j} \, , \quad
\Gamma^t_{it} = \Gamma^t_{ti} = \Gamma^i_{tt} = \Gamma^i_{jk} = 0\, ,
\ee
and
\bea
\label{Rs}
&& R_{tt}= -3 \left( \dot H + H^2 - \dot b H \right)\, , \quad
R_{ij} = \left( \dot H + 3H^2 - \dot b H \right) a^2 \e^{-2b} \delta_{ij}
\, , \quad
R_{it} = R_{ti} = 0 \, , \nn
&& R = 6 \left( \dot H + 2H^2 - \dot b H \right) \e^{-2b}\, ,
\eea
the action (\ref{cos1}) has the following form:
\be
\label{cos2}
S = \int d^4 x a^3 \e^b \left\{
\frac{3}{\kappa^2}\left(\dot H
+ 2 H^2 - \dot b H \right)\e^{-2b} - 27\alpha_0 H^4 \e^{-4b}
\right\}\, .
\ee
We should note $\partial_t \phi = \e^b \sqrt{2U_0}$ and $\partial_i \phi =0$.
It is remarkable that this action is very similar to the one of well-known
$R^2$
gravity.
By the variation of the action with respect to $b$, one obtains an equation
corresponding to the first FRW equation:
\be
\label{cos3}
\frac{3}{\kappa^2} H^2 + 81 \alpha_0 H^4
= \rho_\mathrm{matter}\, .
\ee
Here $\rho_\mathrm{matter}$ is the energy-density of the matter which was
not explicitly written before.
One puts $b=0$ after the variation over $b$.
By the variation over $a$, one also obtains the equation
corresponding to the second FRW equation, which can be evaluated by
using the first FRW equation and matter conservation law.
We should note that the equation (\ref{cos3}) is the first order
differential equation with respect to the scale factor $a(t)$, which should be
distinguished from the usual higher derivative gravity like $F(R)$ gravity,
where the equation corresponding to the first FRW equation is the third order
differential equation with respect to $a(t)$.
Note that $H=0$, which expresses the flat solution, is a trivial solution of
(\ref{cos3}) when there is no matter.
We also note that there is a de Sitter solution, where $H$ is a constant,
given by
\be
\label{coss1}
H^2 = - \frac{1}{27 \alpha_0 \kappa^2}\, ,
\ee
which may express the inflation in the early universe.
The existence of the de Sitter solution in (\ref{coss1}) requires
$\alpha_0<0$, which does not always conflict
with the condition to avoid the tachyon in (\ref{tachyon}).
Now we only kept the term with $n=0$
in the action (\ref{cosppert6}), just for simplicity. If we include the higher
term with $n>0$, the de Sitter solution might become unstable and there can
occur the
instable inflationary solution.

\section{Discussion}

In summary, we formulated covariant higher derivative gravity
with Lagrange multiplier constraint and scalar projectors.
It is demonstrated that such theory admits flat space solution.
Its gauge-fixing formulation is fully developed.
The study of spectrum shows that the only propagating mode is (higher
derivative) graviton, while scalar and vector modes do not propagate.
Eventually, scalar and vector modes correspond to composite states at any
perturbative level.

Furthermore, we show that Lorentz symmetry breaking in the theory under
discussion is spontaneous. The equivalent formulation in terms of
vector fields is developed. The preliminary study of FRW cosmology indicates to
the possibility of inflationary universe solution.
It is interesting that first FRW equation in the theory turns out to be the
first order differential equation which is quite unusual for higher derivative
gravity which normally leads to third order differential equation with respect
to scale factor.
This may indicate to presence of some hidden symmetry in the higher
derivative gravity under consideration.

\section*{Acknowledgments \label{VI}}

We are grateful to I.~Antoniadis for helpful remarks.
This research has been supported in part
by MEC (Spain) project FIS2006-02842 and AGAUR(Catalonia) 2009SGR-994 (SDO), by
Czech Min. of Education under Contract N.MSM 0021622409 (JK),
by Global COE Program of Nagoya University (G07)
provided by the Ministry of Education, Culture, Sports, Science \&
Technology
and by the JSPS Grant-in-Aid for Scientific Research (S) \# 22224003 (SN).


\begin{thebibliography}{99}

%\cite{Nojiri:2010wj}
\bibitem{Nojiri:2010wj}
S.~Nojiri, S.~D.~Odintsov,
%``Unified cosmic history in modified gravity: from F(R) theory to Lorentz
%non-invariant models,''
[arXiv:1011.0544 [gr-qc]], to appear in Phys.\ Rept.; \\
%\cite{Nojiri:2006ri}
%\bibitem{Nojiri:2006ri}
S.~Nojiri and S.~D.~Odintsov,
%``Introduction to modified gravity and gravitational alternative for dark
%energy,''
eConf {\bf C0602061}, 06 (2006)
[Int.\ J.\ Geom.\ Meth.\ Mod.\ Phys.\  {\bf 4}, 115 (2007)]
[arXiv:hep-th/0601213].
%%CITATION = 00436,4,115;%%

\bibitem{CapozzielloFaraoni}
S.~Capozziello and V.~Faraoni,
``Beyond Einstein Gravity'',
{\it Springer, (2010) 428 p}

%\cite{Buchbinder:1992rb}
\bibitem{Buchbinder:1992rb}
I.~L.~Buchbinder, S.~D.~Odintsov and I.~L.~Shapiro,
``Effective action in quantum gravity'',
%\href{http://www.slac.stanford.edu/spires/find/hep/www?irn=2762668}{SPIRES
%entry}
{\it Bristol, UK: IOP (1992) 413 p}

%\cite{Horava:2009uw}
\bibitem{Horava:2009uw}
P.~Horava,
%``Quantum Gravity at a Lifshitz Point,''
Phys.\ Rev.\ D {\bf 79}, 084008 (2009)
[arXiv:0901.3775 [hep-th]].
%%CITATION = PHRVA,D79,084008;%%

%\cite{Nojiri:2009th}
\bibitem{Nojiri:2009th}
S.~Nojiri, S.~D.~Odintsov,
%``Covariant Horava-like renormalizable gravity and its FRW cosmology,''
Phys.\ Rev.\ {\bf D81}, 043001 (2010),
[arXiv:0905.4213 [hep-th]];
%``A proposal for covariant renormalizable field theory of gravity,''
Phys.\ Lett.\ {\bf B691}, 60-64 (2010),
[arXiv:1004.3613 [hep-th]];
%``Covariant power-counting renormalizable gravity: Lorentz symmetry
%breaking and accelerating early-time FRW universe,''
[arXiv:1007.4856 [hep-th]]; \\
%\cite{Chaichian:2011sx}
%\bibitem{Chaichian:2011sx}
M.~Chaichian, M.~Oksanen and A.~Tureanu,
%``ADM representation and Hamiltonian analysis of covariant renormalizable
%gravity,''
arXiv:1101.2843 [gr-qc]; \\
%%CITATION = ARXIV:1101.2843;%%
%\cite{Cognola:2010by}
%\bibitem{Cognola:2010by}
G.~Cognola, E.~Elizalde, L.~Sebastiani and S.~Zerbini,
%``Black hole and de Sitter solutions in a covariant renormalizable field
%theory of gravity,''
Phys.\ Rev.\ D {\bf 83}, 063003 (2011)
[arXiv:1007.4676 [hep-th]].
%%CITATION = PHRVA,D83,063003;%%

\bibitem{extrascalar}
%\cite{Charmousis:2009tc}
%\bibitem{Charmousis:2009tc}
C.~Charmousis, G.~Niz, A.~Padilla and P.~M.~Saffin,
%``Strong coupling in Horava gravity,''
JHEP {\bf 0908}, 070 (2009)
[arXiv:0905.2579 [hep-th]]; \\
%%CITATION = JHEPA,0908,070;%%
%\cite{Li:2009bg}
%\bibitem{Li:2009bg}
M.~Li and Y.~Pang,
%``A Trouble with Horava-Lifshitz Gravity,''
JHEP {\bf 0908}, 015 (2009)
[arXiv:0905.2751 [hep-th]]; \\
%%CITATION = JHEPA,0908,015;%%
%\cite{Blas:2009qj}
%\bibitem{Blas:2009qj}
D.~Blas, O.~Pujolas and S.~Sibiryakov,
%``Consistent Extension of Horava Gravity,''
Phys.\ Rev.\ Lett.\  {\bf 104}, 181302 (2010)
[arXiv:0909.3525 [hep-th]].
%%CITATION = PRLTA,104,181302;%%

\bibitem{vikman}
%\cite{Lim:2010yk}
%\bibitem{Lim:2010yk}
E.~A.~Lim, I.~Sawicki and A.~Vikman,
%``Dust of Dark Energy,''
JCAP {\bf 1005}, 012 (2010)
[arXiv:1003.5751 [astro-ph.CO]]; \\
%%CITATION = JCAPA,1005,012;%%
%\cite{Gao:2010gj}
%\bibitem{Gao:2010gj}
C.~Gao, Y.~Gong, X.~Wang and X.~Chen,
%``Cosmological models with Lagrange Multiplier Field,''
arXiv:1003.6056 [astro-ph.CO]; \\
%%CITATION = ARXIV:1003.6056;%%
%\cite{Capozziello:2010uv}
%\bibitem{Capozziello:2010uv}
S.~Capozziello, J.~Matsumoto, S.~Nojiri and S.~D.~Odintsov,
%``Dark energy from modified gravity with Lagrange multipliers,''
Phys.\ Lett.\ B {\bf 693} (2010) 198
[arXiv:1004.3691 [hep-th]]; \\
%%CITATION = PHLTA,B693,198;%%
%\cite{Cai:2010zma}
%\bibitem{Cai:2010zma}
Y.~F.~Cai and E.~N.~Saridakis,
%``Cyclic cosmology from Lagrange-multiplier modified gravity,''
Class.\ Quant.\ Grav.\ {\bf 28}, 035010 (2011)
[arXiv:1007.3204 [astro-ph.CO]]; \\
%%CITATION = CQGRD,28,035010;%%
%\cite{Kluson:2011xx}
%\bibitem{Kluson:2011xx}
J.~Kluson,
%``Lagrange Multiplier Modified Horava-Lifshitz Gravity,''
arXiv:1101.5880 [hep-th].
%%CITATION = ARXIV:1101.5880;%%

%\cite{Kluson:2010za}
\bibitem{Kluson:2010za}
J.~Kluson, S.~Nojiri, S.~D.~Odintsov and D.~Saez-Gomez,
%``U(1) Invariant F(R)$ Horava-Lifshitz Gravity,''
arXiv:1012.0473 [hep-th].
%%CITATION = ARXIV:1012.0473;%%

\end{thebibliography}
\end{document}